# What Causes High Resistivity in CdTe

Koushik Biswas and Mao-Hua Du

Materials Science & Technology Division and Center for Radiation Detection Materials

and Systems, Oak Ridge National Laboratory, Oak Ridge, TN 37831, USA

CdTe can be made semi-insulating by shallow donor doping. This is routinely done to obtain high resistivity in CdTe-based radiation detectors. However, it is widely believed that the high resistivity in CdTe is due to the Fermi level pinning by native deep donors. The model based on shallow donor compensation of native acceptors was dismissed based on the assumption that it is practically impossible to control the shallow donor doping level so precisely that the free carrier density can be brought below the desired value suitable for radiation detection applications. In this paper, we present our calculations on carrier statistics and energetics of shallow donors and native defects in CdTe. Our results show that the shallow donor can be used to reliably obtain high resistivity in CdTe. Since radiation detection applications require both high resistivity and good carrier transport, one should generally use shallow donors and shallow acceptors for carrier compensation and avoid deep centers that are effective carrier traps.

# I. Introduction

CdTe and its alloys, such as $Cd_{1-x}Zn_xTe$ (CZT), with x = 0.1-0.2, have been extensively studied for their potential applications in room temperature radiation detection.[1] A good detector material must have high resistivity (> $10^9$ Ωcm), which reduces dark current and device noise, and large μτ product (mobility-lifetime product), which ensures that radiation-generated carriers can travel through a large volume of the detector material to be collected by electrodes. The free carrier density in CdTe needs to be controlled below $10^8$ $cm^{-3}$ for radiation detection applications. However, the impurity concentration in CdTe is typically ~$10^{15}$ - $10^{16}$ $cm^{-3}$. Obviously, the high resistivity is made possible by compensation among various native defects and impurities. Meanwhile, these native defects and impurities also affect carrier transport by scattering and trapping free charge carriers. Thus, a proper impurity and defect management is essential for obtaining both high resistivity and large μτ product.

The growth of CdTe or CZT from the melt is typically under Te-rich conditions.[2,3] The higher Cd vapor pressure than that of $Te_2$ causes the loss of Cd during the crystal growth. Therefore, the growth condition is Te-rich even if the starting material is stoichiometric. The resulting undoped CdTe is usually *p*-type with low resistivity, due primarily to the abundance of Cd vacancies ($V_{Cd}$), which are acceptors.[4] To grow semi-insulating CdTe and CZT, one usually needs to use high pressure Bridgman technique,[5,6] which suppresses the loss of Cd during growth, or to dope CdTe with shallow donors,[5,7,8,9,10,11] such as Cl, In, and Al. Although the shallow donor doping is proven to be a very effective approach to obtain high resistivity in CdTe and CZT, it is widely believed that the high resistivity is due to the Fermi level pinning by a deep donor level induced by a native defect, provided that deep donor concentration, $N_{dd}$, exceeds the difference of shallow acceptor, $N_{sa}$, and shallow donor concentrations, $N_{sd}$ ($N_{dd} > N_{sa} - N_{sd}$).[12, 13, 14] This deep donor is usually assumed to be Te antisite (Te$_{Cd}$).[15, 16, 17, 18] The rationale of the deep donor mediated carrier compensation model is that it is practically impossible to control the shallow donor concentration to almost exactly compensate the native and impurity acceptors such that the free carrier density can be less than $10^8$ cm$^{-3}$.[13, 19] However, the deep donor mediated carrier compensation model has several serious flaws as we discuss below.

(1) We contend that it is entirely possible to compensate excess acceptors in CdTe reliably using shallow donors and obtain a free carrier density less than $10^8$ cm$^{-3}$. Previous calculations on carrier statistics and resistivity in CdTe conclude that high resistivity in CdTe is attainable only by deep donor doping.[13, 22] However, these calculations are all based on the incorrect assumption that the native acceptor concentration is a constant and independent of the Fermi level. This assumption has been invoked numerous times for calculations of the carrier density, the level of the hypothetical deep donor, resistivity, etc. These calculations all missed an important fact that the formation energy of an acceptor defect decreases with rising Fermi level. Therefore, a higher doping level of the shallow donors in CdTe should be accompanied by an automatic increase of the native acceptor concentration that partially compensates the rise of the shallow donor concentration. This results in a smooth and relatively slow change of the Fermi level with the rise of the shallow donor concentration. The calculations that take into account the dependence of both donors and acceptors on the Fermi level show that the high resistivity (> $10^9$ Ωcm) can be maintained over a large range of the shallow donor concentration. This range is at least in the order of ~$10^{16}$ - $10^{17}$ cm$^{-3}$ and can even be up to the solid solubility of the donor (e.g., >$10^{19}$ cm$^{-3}$ for Cl) as we will demonstrate in Section III-A.

(2) The deep donor model cannot reconcile with the good µτ product in high-quality detector grade CdTe and CZT. It is well known that doping by extrinsic deep donors, such as Ge and Sn, can give rise to high resistivity.[15, 19, 20, 21] However, the µτ product in Ge or Sn doped CdTe and CZT (with the dopant concentration ~ $10^{16}$ cm$^{-3}$) are significantly reduced due to the electron trapping at the deep donor level.[20] There is no apparent reason to believe that other deep donors, such as Te$_{Cd}$, will be benign in terms of electron trapping. The fact that the detector-grade highly resistive CdTe and CZT have good µτ products suggests that the ionized deep donor concentration is low. It has been estimated that the concentration of a hypothetical midgap donor should be very low (in the order of $10^{11}$ cm$^{-3}$) based on the measured electron µτ product and the capture cross section of the deep-level defects in the high-quality detector grade CZT.[22] Such a small deep donor concentration cannot possibly control the conductivity of CdTe and CZT, which have much higher concentration of native defects and residual impurities (~$10^{15}$-$10^{16}$ cm$^{-3}$). In semi-insulating CdTe and CZT intentionally doped with shallow donors, the shallow donor concentration can be very high, e.g., [Cl] ~ $10^{19}$ cm$^{-3}$ in CdTe:Cl and [In] ~ $10^{16}$-$10^{17}$ cm$^{-3}$ in CdTe:In.[5, 7, 8, 9] It should be obvious that the shallow donors and shallow acceptors dominate the carrier compensation in heavily doped CdTe and CZT.

(3) The deep native donor is usually assumed to be Te$_{Cd}$.[15,16,17,18] However, there is no experimental proof that Te$_{Cd}$ does introduce a deep donor level near midgap. Many efforts have been devoted to the measurement of the deep levels in CdTe and CZT[17, 23, 24, [25] and it is often suggested that some observed deep levels are responsible for the high resistivity. However, the concentration of these deep level defects is unknown and there is no proof that they control the conductivity. A recent study shows that some deep levels may be induced by dislocations.[26]

It can be seen from the discussion above that the deep-donor-mediated carrier compensation model in CdTe is not well justified. Alternative explanations to the high resistivity in CdTe have been pursued. It has been suggested that the O-H complex may be important in carrier compensation in CdTe.[27, 28] However, in heavily doped CdTe, such as CdTe:Cl with [Cl]~ $10^{19}$ cm$^{-3}$, the role that the O-H complex may play should be limited due to the much lower oxygen concentration ~$10^{16}$ cm$^{-3}$.[17, 29]

First-principles calculations based on density functional theory (DFT) within local density approximation (LDA) have been used to study the defects and impurities in CdTe.[30, 25, 27, 28, 31] It is shown in Ref. 30 that Te$_{Cd}$ induces a (+/0) and a (2+/+) transition levels at $E_c$ – 0.34 eV and $E_c$ – 0.59 eV, respectively, in the upper half of the band gap ($E_c$ is the energy of the conduction band edge). More recent calculations[27, 28, 31] show that there should be a stronger Jahn-Teller distortion at neutral Te$_{Cd}$ than predicted in Ref. 30. This results in a much more stable neutral Te$_{Cd}$ and consequently lower donor level. The (2+/0) transition level of Te$_{Cd}$ is calculated to be at $E_v$ + 0.35 eV ($E_v$ is the energy of the valence band edge).[27, 28] A combined experimental and theoretical study shows that the observed midgap levels may be deep acceptor levels caused by Te$_{Cd}$ –$V_{Cd}$ complex.[25] Our present calculations show that the Te$_{Cd}$ –$V_{Cd}$ complex indeed induces deep acceptor levels near midgap but it has a high formation energy (as will be shown in Section III-D) and thus cannot play a significant role in carrier compensation.

In this paper, we present our calculations on carrier statistics and resistivity of CdTe. Hybrid functional calculations[32] were employed to calculate the formation energies of native defects and shallow donors. The calculated formation energies were used in the carrier statistics calculations. The hybrid functional calculations, which can partially remove the self-interaction error and correct the band gap, have recently been applied to defect calculations and have generally shown improvement in structural, electronic, dielectric, and defect properties in semiconductors.[33, 34, 35, 36, 37, 38, 39, 40, 41, 42, 43, 44, 45, 46, 47, 48] Our results show that the shallow donor can be used reliably to obtain high resistivity in CdTe. Reasonable fluctuations in the doping level during the crystal growth can be tolerated.

## II. Methods

### A. Computational Details

Density functional calculations were performed to study the native defects and shallow donor impurities in CdTe. Hybrid functionals,[32] as implemented in the VASP codes,[49] were used to calculate defect formation energies. The fraction of Hartree-Fock exchange was adjusted to 16.5% to correct the band gap to 1.55 eV, which is in good agreement with the experimental value of 1.61 eV[50]. The electron-ion interactions were described using projector augmented wave potentials.[51] The wavefunctions were expanded in a plane-wave basis with cutoff energy of 275 eV. The Cd $d$ electrons were treated as valence electrons. The lattice constant of CdTe is calculated to be 6.59 Å, in good agreement with the experimental lattice constant of 6.477 Å.[52] The bulk host material was simulated using a 64 atom cubic supercell and the Brillouin-zone sampling was performed using a 2×2×2 Γ-centered $k$-mesh. The force on each atom was minimized to be less than 0.05 eV/Å. Corrections to image charges in supercell calculations were performed following Ref. 53. We have performed convergence tests using Perdew-Burke-Ernzerhof (PBE)[54] functionals. The formation energy of $V_{Cd}^{2-}$ was calculated using a denser $k$-mesh (4×4×4), a higher cutoff energy (400 eV), and a larger supercell size (216-atom cell). In each case, the change in formation energy is well below 0.05 eV.

The defect formation energy of a defect in charge state $q$ is given by,

$$\Delta H = (E_{Defect} - E_{host}) - \sum_i n_i(\mu_i + \mu_i^{ref}) + q(\varepsilon_{VBM} + \varepsilon_f) \qquad (1)$$

where $E_{Defect}$ and $E_{host}$ are the total energies of the defect-containing and the host (i.e. defect-free) supercells. $n_i$ is the difference in the number of atoms for the $i$th atomic species between the defect-containing and defect-free supercells. $\mu_i$ is a relative chemical potential for the $i$th atomic species, referenced to $\mu_i^{ref}$. For Cd and Te, $\mu_{Cd}^{ref}$ and $\mu_{Te}^{ref}$ are the chemical potentials in bulk Cd and bulk Te, respectively. The third term in eq.(1) represents the change in energy due to exchange of electrons or holes with the respective carrier reservoirs. $\varepsilon_{VBM}$ is the energy of the valence band maximum (VBM) in the host system and $\varepsilon_f$ is the Fermi energy relative to the VBM.

The defect concentration at thermal equilibrium can be evaluated using

$$N = N_{site}\exp(-\Delta H/kT), \qquad (2)$$

where $N_{site}$ is the number of available sites for defect formation in the crystal, $\Delta H$ is the defect formation energy defined in Eq. 1, $k$ is the Boltzmann's constant, and $T$ is the temperature. The transition level, $\varepsilon(q/q')$, of a defect is the value of $\varepsilon_f$ at which the formation energies in charge states $q$ and $q'$ are identical.

The calculated heat of formation for CdTe, $\Delta H$(CdTe), is -1.13 eV, in reasonable agreement with the experimental value of –0.96 eV.[55] The stability condition for the formation of CdTe under equilibrium is given by $\mu_{Cd} + \mu_{Te} = \Delta H(\text{CdTe}) = -1.13$ eV. The following conditions are also applied to prevent the formation of various competing phases (i.e., $CdCl_2$, $TeCl_4$, $Te_3Cl_2$, $In_2Te_3$, and $In_2Te_5$) when Cl and In donor impurities are introduced into CdTe:

$$\mu_{Cd} + 2\mu_{Cl} \leq \Delta H(\text{CdCl}_2) = -3.58 \text{ eV}$$

$$\mu_{Te} + 4\mu_{Cl} \leq \Delta H(\text{TeCl}_4) = -3.14 \text{ eV}$$

$$3\mu_{Te} + 2\mu_{Cl} \leq \Delta H(\text{Te}_3\text{Cl}_2) = -1.56 \text{ eV} \quad (3)$$

$$2\mu_{In} + 3\mu_{Te} \leq \Delta H(\text{In}_2\text{Te}_3) = -1.66 \text{ eV}$$

$$2\mu_{In} + 5\mu_{Te} \leq \Delta H(\text{In}_2\text{Te}_5) = -1.99 \text{ eV}$$

Here, $\mu_{Cl}$ and $\mu_{In}$ are the relative Cl and In chemical potentials referenced to those for $Cl_2$ molecule and the In metal, respectively.

### B. Carrier Statistics and Resistivity

The free carrier densities of electrons ($n$) and holes ($p$) as well as the Fermi level can be obtained by self-consistently solving the Eqs. 4-6 below.

$$p + \sum_i \sum_j N_{D,i}(q_j) = n + \sum_i \sum_j N_{A,i}(q_j) \quad (4)$$

$$n = N_C e^{-(E_g - \varepsilon_f)/k_B T} \quad (5)$$

$$p = N_V e^{-\varepsilon_f/k_B T} \qquad (6)$$

Here, $N_{D,i}(q_j)$ ($N_{A,i}(q_j)$) is the density of the ionized donor (acceptor) of $i$th type and charge state $q_j$. Both of them are functions of $q\varepsilon_f$ as shown in Eq. (1). $E_g$ is the band gap of CdTe. $N_C$ and $N_V$ are the effective densities of states of conduction and valence bands, respectively. They can be evaluated by using $N_C = 2(2\pi m_e k_B T/h^2)^{3/2}$ and $N_V = 2(2\pi m_h k_B T/h^2)^{3/2}$, where $m_e$ and $m_h$ are the effective masses of the electron and the hole, respectively, and $h$ is the Planck constant. The resistivity can be further calculated using the equation

$$\rho = \frac{1}{e(p\mu_p + n\mu_n)}, \qquad (7)$$

where $\mu_n$ ($\mu_p$) is the mobility of the electron (hole). We use the following values for parameters in Eqs. 2, 5-7: $N_{site} = 1.47 \times 10^{22}$ cm$^{-3}$, $m_e = 0.11\ m_0$, $m_h = 0.73\ m_0$, $\mu_n = 1000$ cm$^2$/Vs, $\mu_p = 80$ cm$^2$/Vs, $E_g = 1.55$ eV. The band gap is calculated by the hybrid functional calculation.

A large CdTe single crystal is usually grown from the melt and subsequently cooled down to the room temperature. The atoms and electrons in the crystal should attempt to equilibrate with their respective reservoirs when CdTe is cooled from the growth temperature to the room temperature. The atoms will freeze in the lattice relatively quickly when the crystal is cooled whereas the localized charges in defects and impurities and the free carriers may equilibrate at the room temperature. Therefore, the concentrations of the native defects and impurities are calculated using the growth temperature typically applied to CdTe, i.e., $T = T_{growth} = 1373$ K, while the carrier density is calculated using the room temperature, i.e., $T = T_{room} = 300$ K. The total electric charge for the shallow acceptor/donor is simply the local charge on the shallow acceptor/donor multiplied by the concentration of the shallow acceptor/donor calculated at $T = 1373$ K. Although the total number of a donor or an acceptor is determined at $T = 1373$ K, its charge may re-equilibrate at $T = 300$ K if the donor or the acceptor level is not far from the Fermi level. This will change the ratio of the number of the ionized donors/acceptors to that of the neutral ones. In the carrier statistics calculations shown in Section III-A, several different types of donors and acceptors are involved, i.e., a shallow single-electron donor, a deep double-electron donor, a shallow single-electron acceptor, and a shallow double-electron acceptor. We assume that all the shallow donors and the shallow acceptors are fully ionized at room temperature since we are only interested in CdTe with its Fermi level in the middle region of the band gap. For the deep double-electron donor, if the deep donor level ($\varepsilon_d$) is fully occupied, the deep donor would be charge neutral. Thus, the total charge of all the deep double-electron donors ($Q_{DD}$) is determined by the number of holes on the donor level at $T = 300$ K:

$$Q_{DD} = 2eN_{DD}\left(1 - \frac{1}{e^{(\varepsilon_d - \varepsilon_f)/k_B T} + 1}\right) = \frac{2eN_{DD}}{1 + e^{(\varepsilon_f - \varepsilon_d)/k_B T}}, \tag{8}$$

where $N_{DD}$ is the number of the deep donors. Here, we neglect any electronic or structural relaxation upon the charge state change for the deep donor. Thus, the deep donor level position is independent on its occupation.

## III. Results and Discussion

### A. Compensation in CdTe: Carrier Statistics and Resistivity

We first consider the carrier compensation between a shallow double-electron acceptor and a shallow single-electron donor. This simulates the case of shallow donor ($D^+$) doping of CdTe where the Cd vacancy ($V_{Cd}^{2-}$) is abundant. The formation energy of $V_{Cd}^{2-}$ is calculated by hybrid functional calculations, i.e., $\Delta H(V_{Cd}^{2-}) = 3.063$ eV $- 2\varepsilon_f$ (following Eq. 1). The Cd vacancy concentration, [$V_{Cd}^{2-}$], is calculated at 1373 K using Eq. 2. When the Fermi level is at midgap ($E_v + 0.775$ eV), [$V_{Cd}^{2-}$] is $4.11 \times 10^{16}$ cm$^{-3}$. This is in contrast to previous calculations where [$V_{Cd}^{2-}$] is fixed.[13] The shallow donor concentration, [$D^+$], is varied from $1.0 \times 10^{14}$ cm$^{-3}$ to $3.0 \times 10^{19}$ cm$^{-3}$ in the calculation.

The calculated resistivity and the Fermi level as functions of [$D^+$] are shown in Fig. 1. It can be seen that the resistivity stays above $10^9$ Ωcm when [$D^+$] increases from ~$1.0 \times 10^{16}$ cm$^{-3}$ to ~$5.0 \times 10^{17}$ cm$^{-3}$. This means that the experimental control of [$D^+$] within a few times of $10^{16}$ cm$^{-3}$ is sufficient for obtaining high resistivity reliably. If [$V_{Cd}^{2-}$] is fixed as done in the past, the resistivity curve would be like a δ-function centered near [$D^+$] = 2[$V_{Cd}^{2-}$]. This is clearly an artifact due to the incorrect assumption made in the previous calculations.

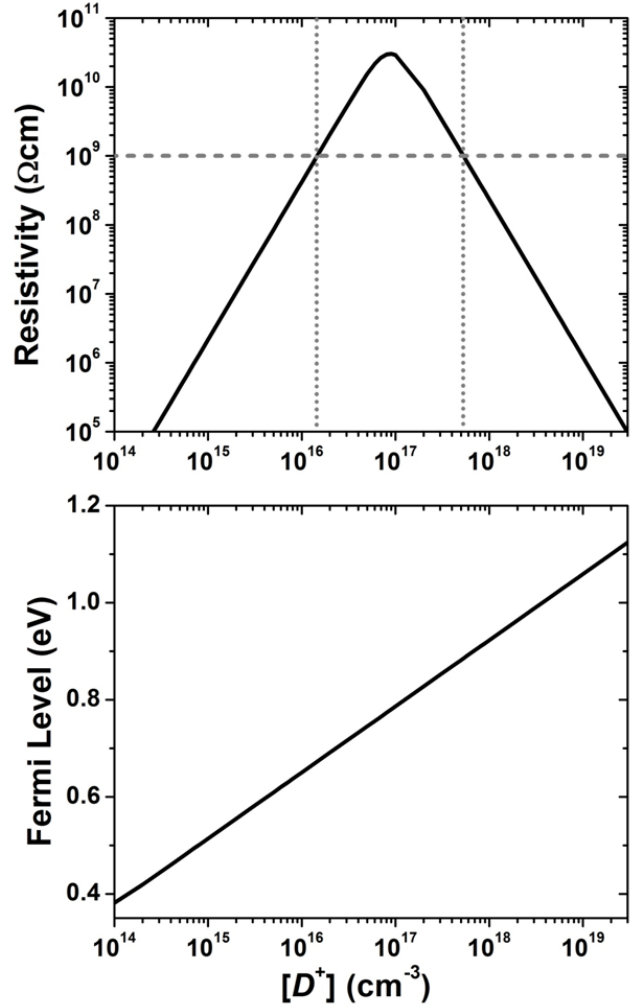

**FIGURE 1. Calculated resistivity (a) and the Fermi level position (b) of CdTe as functions of the shallow donor concentration $[D^+]$. A shallow single-electron donor ($D^+$) and a shallow double-electron acceptor ($V_{Cd}^{2-}$) are considered in the calculation. A resistivity higher than $10^9$ Ωcm is usually required for radiation detector materials.**

Next, we turn to the deep donor. The shallow single-electron donor, $D^+$, in the calculation is now replaced by a deep double-electron donor, $DD$, e.g., $Ge_{Cd}$, $Sn_{Cd}$. Three different deep donor levels are considered in the calculations: $E_v + 0.6$ eV, $E_v + 0.75$ eV, and $E_v + 1.3$ eV. The resistivity of CdTe is calculated with the deep donor concentration, $[DD]$, being varied from $1.0 \times 10^{14}$ cm$^{-3}$ to $3.0 \times 10^{19}$ cm$^{-3}$ in the calculation. Fig. 2(a) shows that the high resistivity ($> 10^9$ $\Omega$cm) is obtained in all three cases. Fig. 2(b) shows that the Fermi level rises with the deep donor concentration, [DD], but at a much slower pace when the Fermi level is above the donor level. The presence of a deep donor level reduces the fraction of the ionized donors and thus slows down the rise of the Fermi level in response to the increasing [DD]. The deeper the donor level is, the slower $\varepsilon_f$ would rise with [DD]. As a result, a deeper donor usually has a larger range of [DD] that gives high resistivity as manifested by Fig. 2(a). However, there could be exceptions, i.e., if the deep donor level is much lower than the midgap, $\varepsilon_f$ may not be able to rise to the midgap before the donor concentration reaches its solid solubility.

Although the deep donor is very effective in carrier compensation, it is also an efficient carrier trap and thus should be avoided in radiation detectors. On the other hand, the shallow donor can be reliably used to obtain high resistivity in CdTe, as manifested in Fig. 1. As can be seen from Fig. 1(b), the Fermi level rises slowly with $[D^+]$ (linear to $\log[D^+]$), allowing reasonable fluctuations of the donor concentration during the crystal growth without lowering the resistivity significantly. Next, we will show that the Fermi level can further be stabilized over a wide range of $[D^+]$ by the formation of *A* centers and the resulting donor self-compensation.

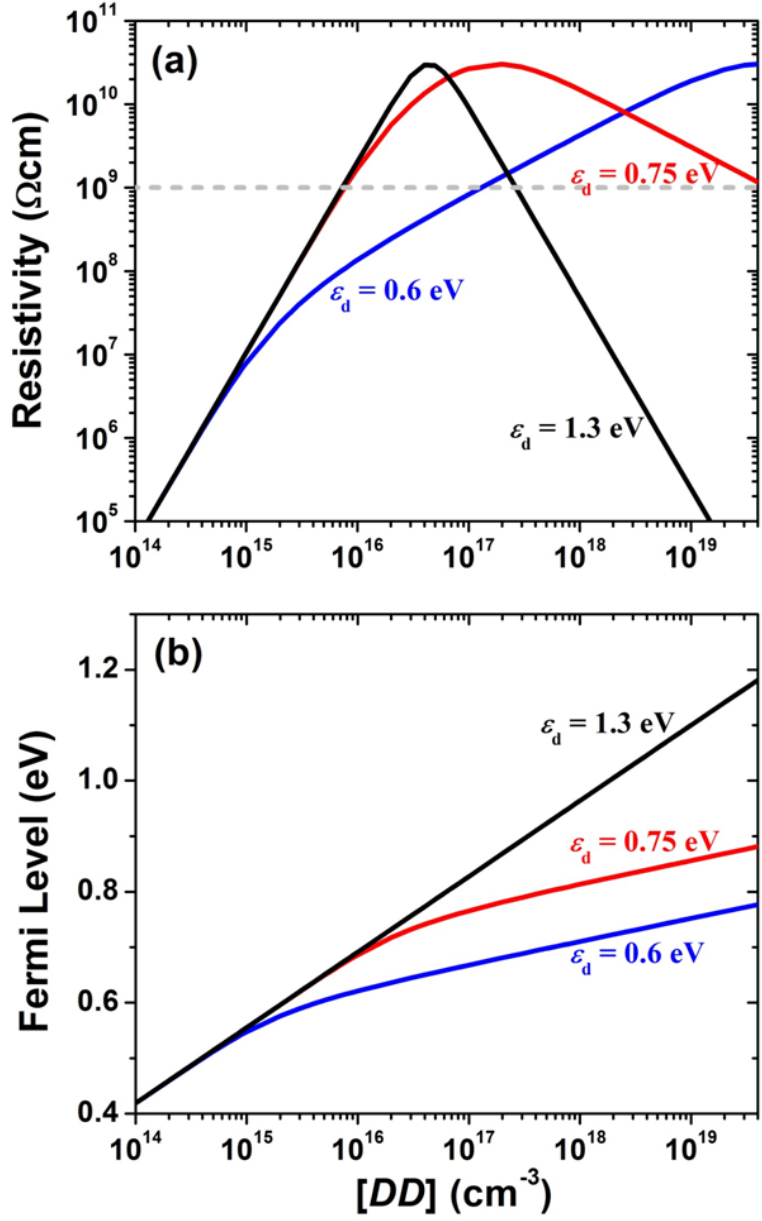

**FIGURE 2.** (Color online) Calculated resistivity (a) and the Fermi level position (b) of CdTe as functions of the deep donor concentration, [*DD*]. A deep double-electron donor (*DD*) and a shallow double-electron acceptor ($V_{Cd}^{2-}$) are considered in the calculation. The results for three different deep donor levels, i.e., $\varepsilon_d$ = 0.6 eV, 0.75 eV, and 1.3 eV, are shown.

A shallow donor, $D^+$, can bind with a $V_{Cd}^{2-}$ to form an *A* center, which is a single-electron acceptor. Increasing $[D^+]$ will induce the formation of more Cd vacancies. Some of them are isolated Cd vacancies and some others are the Cd vacancies bound in the *A* centers. The Cd vacancy concentration, [$V_{Cd}^{2-}$], mentioned in this paper always means the concentration of the isolated Cd vacancy, not including those bound in the *A* centers.

The *A* center concentration, [$A^-$], increases with [$D^+$], and may exceed the concentration of the isolated $V_{Cd}^{2-}$, [$V_{Cd}^{2-}$], at sufficiently high [$D^+$]. When that happens, the Fermi level will no longer change with [$D^+$] due to the donor self-compensation. The exact location where the Fermi level can be stabilized depends on the detailed defect energetics. Generally, the stability of the *A* center increases with its binding energy, $\Delta B$. A larger $\Delta B$ would stabilize the Fermi level at a lower position in the band gap as shown in Fig. 3(b). If $\Delta B$ is relatively small (e.g., $\Delta B$ = 0.6 eV and 1.0 eV in Fig. 3), the *A* center would outnumber $V_{Cd}^{2-}$ only when [$D^+$] is high. Thus, the Fermi level would be stabilized in the *n*-type region. [Note that the Fermi level for $\Delta B$ = 0.6 eV is stabilized at a higher [$D^+$] that is beyond the upper bound of [$D^+$] in Fig. 3] In the case of very small $\Delta B$, such as $\Delta B$ = 0, the Fermi level and the resistivity will essentially respond to the increasing [$D^+$] in the way shown in Fig. 1. On the other hand, if $\Delta B$ is very large (e.g., $\Delta B$ = 1.8 eV in Fig. 3), the Fermi level would be stabilized in the *p*-type region. Our calculations show that, with $\Delta B$ in the range of 1.29 eV - 1.71 eV, the Fermi level is stabilized near the midgap, resulting in high resistivity (> $10^9$ $\Omega$cm) even if [$D^+$] approaches its solid solubility [see, for example, in Fig. 3(a) for $\Delta B$ = 1.3 or 1.6 eV]. However, it should be mentioned that, since the donors scatter the electron carriers, one should obtain the high resistivity using as few shallow donors as possible.

We will show in the next section how the resistivity and the Fermi level shown in this section can be related to the defect energetics.

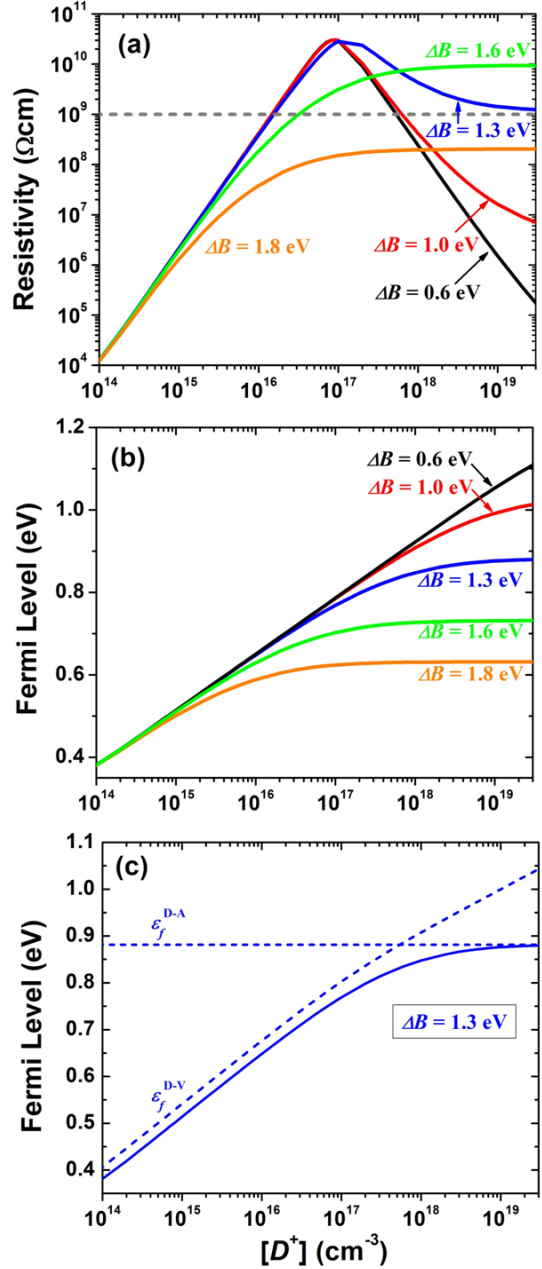

FIGURE 3. (Color online) Calculated resistivity (a) and the Fermi level (b) of CdTe as functions of the shallow donor concentration $[D^+]$. A shallow single-electron donor ($D^+$), a shallow double-electron acceptor ($V_{Cd}^{2-}$), and a shallow single-electron acceptor (the $A$ center) are considered in the calculation. The results for five different binding energies (i.e., $\Delta B$ = 0.6, 1.0 eV, 1.3 eV, 1.6 eV, and 1.8 eV) between the shallow donor ($D^+$) and the Cd vacancy ($V_{Cd}^{2-}$) in the $A$ center are shown in (a) and (b). The Fermi level as a functional of $[D^+]$ for $\Delta B$ = 1.3 eV is also shown in (c) with additional lines for $\varepsilon_f^{D-V}$ and $\varepsilon_f^{D-A}$ shown (see text for details).

## B. Energetics of Donors and Acceptors: General Picture

When $\varepsilon_f$ is near midgap, the free carrier density is negligible compared to the concentrations of the donor and the acceptor. Thus, if we plot the formation energies of $V_{Cd}^{2-}$ and $D^+$ as functions of $\varepsilon_f$, their crossing point ($\varepsilon_f^{D-V}$), where the formation energies of $D^+$ and $V_{Cd}^{2-}$ are the same, should be approximately equal to the equilibrium Fermi level of the system [see, for example, a schematic in Fig. 4(a)]. During the crystal growth of CdTe, if more shallow donors are added, the donor chemical potential would rise and consequently its formation energy would decrease (see Eq. 1). As a result, $\varepsilon_f^{D-V}$, which is approximately the Fermi level pinning point ($\varepsilon_f^{pin}$), moves higher with increasing $[D^+]$, as schematically shown from Fig. 4(a) through Fig. 4(c). Similar to $\varepsilon_f^{D-V}$, $\varepsilon_f^{D-A}$ is the Fermi energy where the formation energy lines of $D^+$ and $A^-$ cross. However, notice that, unlike $\varepsilon_f^{D-V}$, the location of $\varepsilon_f^{D-A}$ does not change with $[D^+]$. This is because an $A$ center contains one shallow donor atom and the formation energies of $D^+$ and $A^-$ decrease simultaneously at the same rate with increasing $[D^+]$, which keeps $\varepsilon_f^{D-A}$ stationary. This is schematically shown from Fig. 4(a) through Fig. 4(c). Note that $\varepsilon_f^{D-V} < \varepsilon_f^{D-A}$ in Fig. 4(a) and $\varepsilon_f^{D-V} = \varepsilon_f^{D-A}$ in Fig. 4(b). Continuing increasing $[D^+]$ will eventually cause $\varepsilon_f^{D-V} > \varepsilon_f^{D-A}$ as shown in Fig. 4(c). Now the $A$ center becomes the dominant acceptor. Since the Fermi level should be pinned by the lowest crossing point between a donor and an acceptor formation energy lines, $\varepsilon_f^{pin} \approx \varepsilon_f^{D-A}$ in Fig. 4(c). Consequently, the Fermi level will no longer rise with $[D^+]$.

Now let us define a range of $\varepsilon_f$, bounded by $\varepsilon_{f1}$ and $\varepsilon_{f2}$ [shaded area in Fig. 4], within which the resistivity of CdTe is sufficiently high for radiation detection purpose. Fig. 4 illustrates the case for $\varepsilon_f^{D-A} > \varepsilon_{f2}$, in which case increasing $[D^+]$ eventually leads to *n*-type CdTe. The electron density will saturate when the Fermi level reaches $\varepsilon_f^{D-A}$.

Fig. 5 is same as Fig. 4 except that the location of $\varepsilon_f^{D-A}$ is now near midgap ($\varepsilon_{f1} < \varepsilon_f^{D-A} < \varepsilon_{f2}$). Since $\varepsilon_f^{pin}$ cannot move above $\varepsilon_f^{D-A}$, the Fermi level is fixed near midgap even if $[D^+]$ reaches its solid solubility. This results in high resistivity over a wide range of $[D^+]$. This scenario is illustrated in Fig. 3(c) as an example. As can be seen in Fig. 3(c), as $[D^+]$ increases, the Fermi level initially tracks $\varepsilon_f^{D-V}$ when $\varepsilon_f^{D-V} < \varepsilon_f^{D-A}$, and then becomes stabilized at $\varepsilon_f^{D-A}$ (which is a constant) when $\varepsilon_f^{D-V} > \varepsilon_f^{D-A}$.

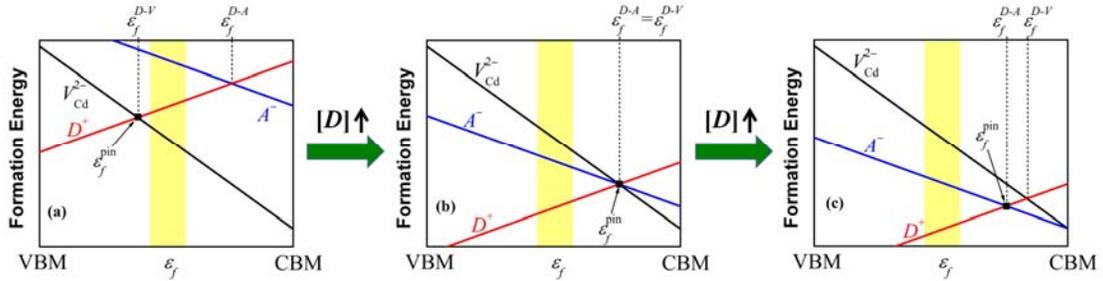

**FIGURE 4. (Color online) Schematic figures for formation energies of a shallow donor ($D^+$), Cd vacancy ($V_{Cd}^{2-}$), and the complex ($V_{Cd}$ -$D$)⁻ (*A* center) in CdTe as functions of the Fermi level ($\varepsilon_f$). The shallow donor concentration, [$D$], increases from (a), (b), to (c). The slope of a formation energy line indicates the charge state of the defect. The shaded region near the middle of the band gap show schematically the lower and upper bounds of the Fermi level (i.e., $\varepsilon_{f1}$ and $\varepsilon_{f2}$), within which the resistivity of CdTe is sufficiently high for radiation detection applications. See text for details.**

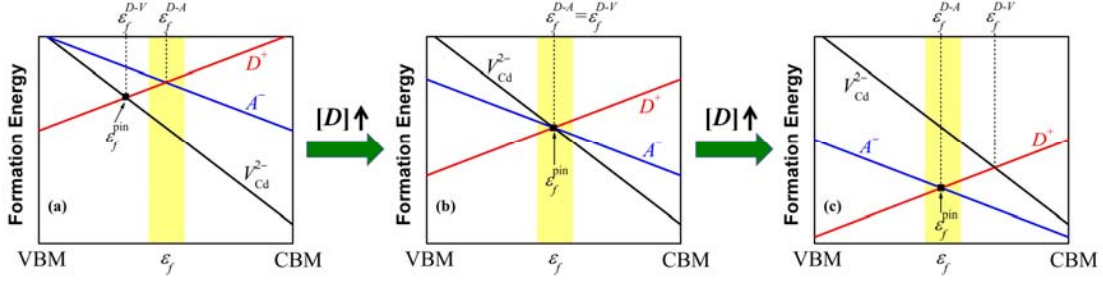

**FIGURE 5. Same as Fig. 4 except that, here, $\varepsilon_f^{D-A}$ is in the midgap region ($\varepsilon_f^{min} < \varepsilon_f^{D-A} < \varepsilon_f^{max}$). See text for details.**

In the case of $\varepsilon_f^{D-A} < \varepsilon_{f1}$, CdTe will remain *p*-type with relatively low resistivity regardless how many shallow donors are introduced into CdTe. This should correspond to the scenario shown in Figs. 3(a) and (b) for $\Delta B = 1.8$ eV.

The different location of $\varepsilon_f^{D-A}$ as shown in Figs. 4 and 5 depends on the binding energy between $D^+$ and $V_{Cd}^{2-}$. The formation energies of $V_{Cd}^{2-}$ and $A^-$ can be written as $\Delta H(V_{Cd}^{2-}) = a - 2\varepsilon_f$ and $\Delta H(A^-) = \Delta H(V_{Cd}^{2-}) + \Delta H(D^+) - \Delta B$, where $a$ is the formation energy of $V_{Cd}^{2-}$ at the VBM, $\Delta H(D^+)$ is the formation energy of the shallow donor, and $\Delta B$ is the *A* center binding energy. It can then be derived that

$$\varepsilon_f^{D-A} = (a - \Delta B)/2. \tag{9}$$

Thus, a stronger binding energy leads to a lower $\varepsilon_f^{D-A}$ with respect to the conduction band minimum (CBM).

The results in Section III-A and the analyses performed in this section show that high resistivity (> $10^9$ Ωcm) can be obtained reliably by the shallow donor doping unless $\varepsilon_f^{D-A}$, which is the upper limit of the Fermi level, is too low in the band gap. As long as $\varepsilon_f^{D-A}$ is near or above the midgap, all that is required for achieving high resistivity is a high enough solid solubility for the shallow donor such that the Fermi level can be raised to the midgap region.

In the case of Fermi level pinning by $\varepsilon_f^{D-V}$, the tolerable uncertainty in the doping level ($\delta N$) is in the order of $10^{16}$ - $10^{17}$ cm$^{-3}$ based on the calculated formation energy of $V_{Cd}^{2-}$ (see Section III-A). Now we will give a general assessment of $\delta N$ without relying on the calculated [$V_{Cd}^{2-}$]. If the shallow donor concentrations are $N_{D1}$ and $N_{D2}$ when $\varepsilon_f = \varepsilon_{f1}$ and $\varepsilon_{f2}$, respectively, it is easy to show that $N_{D2}/N_{D1} = \exp(2\Delta\varepsilon_f/kT)$, where $\Delta\varepsilon_f = \varepsilon_{f2} - \varepsilon_{f1}$. Using a crystal growth temperature of 1375 K and $\Delta\varepsilon_f < 0.275$ eV for CdTe, one finds that $N_{D2}/N_{D1}$ ranges from ~$10^0$ to ~$10^2$. Therefore, an increase in [$D^+$] by nearly two orders of magnitude is needed to move the Fermi level from $\varepsilon_{f1}$ to $\varepsilon_{f2}$. If we tighten the resistivity requirement such that $\Delta\varepsilon_f < 0.15$ eV, $N_{D2}/N_{D1}$ would range from 1 to 12.6, which still requires more than one order of magnitude increase in [$D^+$] to move the Fermi level from $\varepsilon_{f1}$ to $\varepsilon_{f2}$. Therefore, the tolerable uncertainty in the shallow donor concentration can be quite large. Since the shallow donor concentration is approximately twice of [$V_{Cd}^{2-}$] when the Fermi level is pinned by $\varepsilon_f^{D-V}$, a larger [$V_{Cd}^{2-}$] under thermal equilibrium should lead to a larger tolerable uncertainty in the shallow donor concentration. Note that the above discussion is based on the scenario where the Fermi level is pinned by $\varepsilon_f^{D-V}$. If the Fermi level can be pinned by $\varepsilon_f^{D-A}$ near midgap, the allowed shallow doping range would be much larger, extending to the solid solubility of the shallow donor.

## C. Energetics of Shallow donors: In and Cl

The reported In concentrations in In-doped semi-insulating CZT are typically in the order of $10^{16}$ or $10^{17}$ cm$^{-3}$, e.g., [In] = $2.2 \times 10^{16}$ - $2.2 \times 10^{17}$ cm$^{-3}$ ($\rho$ = 3.4 - 4.1 $\times 10^{10}$ $\Omega$cm) (Ref. 9), [In] ~ $10^{17}$ cm$^{-3}$ ($\rho$ = 0.2 - 5 $\times 10^{10}$ $\Omega$cm) (Ref.10), and [In] ~ $1 \times 10^{17}$ cm$^{-3}$ ($\rho$ = 3.3 - 3.4 $\times 10^{10}$ $\Omega$cm) (Ref. 11). These doping levels are in good agreement with the calculated shallow donor concentrations that are needed for high resistivity (see Fig. 1). A much wider range of Cl concentration in Cl-doped semi-insulating CdTe and its alloys has been reported, e.g., [Cl] = $2 \times 10^{19}$ cm$^{-3}$ in CdTe ($\rho = 1.0 \times 10^9$ $\Omega$cm) (Ref. 7), 1-2.7$\times 10^{19}$ cm$^{-3}$ in CdTe ($\rho = 4 \times 10^9$ - $1 \times 10^{10}$ $\Omega$cm) (Ref. 8), and [Cl] = $5 \times 10^{17}$ cm$^{-3}$ in CdTe$_{0.9}$Se$_{0.1}$:Cl ($\rho = 4.5 \times 10^9$ $\Omega$cm) (Ref. 56). A shallow donor concentration as high as $10^{19}$ cm$^{-3}$ would clearly make CdTe *n*-type (see Fig. 1) unless the shallow donor and its *A* center can pin the Fermi level near midgap. This requires a proper *A* center binding energy as discussed in Section III-B. The experimental results on CdTe:In and CdTe:Cl may be explained if CdTe:In resembles the doping scenario described by results in Fig. 3 (for $\Delta B$ = 0.6 or 1.0 eV) and Fig. 4 while CdTe:Cl resembles those shown in Fig. 3 (for $\Delta B$ = 1.3 or 1.6 eV) and Fig. 5. The combination of the experimental results and the calculations and analyses shown in Section III-A and III-B suggest that (1) in semi-insulating CdTe:In, the Fermi level is pinned by $In^+_{Cd}$ and $V^{2-}_{Cd}$; (2) in semi-insulating CdTe:Cl, the Fermi level is pinned by $Cl^+_{Te}$ and $V^{2-}_{Cd}$ if [Cl] is relatively low (e.g., ~$10^{16}$ cm$^{-3}$) and by $Cl^+_{Te}$ and its *A* center when [Cl] is high (e.g., ~$10^{19}$ cm$^{-3}$); (3) The binding energy of Cl-*A* center should be larger than that of In-*A* center (which may be expected as Cl$_{Te}$ is the nearest neighbor while In$_{Cd}$ is the second-nearest neighbor to the Cd vacancy), and $\varepsilon_f^{D-A}$ for CdTe:Cl is near midgap and lower than that for CdTe:In; (4) increasing In or Cl donor density to its solid solubility will result in *n*-type CdTe:In and semi-insulating CdTe:Cl, respectively.

We have shown in Section III-A that the high resistivity can always be obtained by the shallow donor doping, provided that the solid solubility of the shallow donor is high enough to raise the Fermi level to midgap and the *A*-center binding energy has a proper value so that $\varepsilon_f^{D-A}$ is sufficiently high (near or above the midgap). Figures 6 and 7 show the formation energies of In and Cl related defects in CdTe, respectively. Maximum In and Cl chemical potentials allowed by Eqs. 3 are used. Under the Te-rich and shallow-donor-rich limits [Fig. 6(a) and 7(a)], the In and Cl solid solubilities (including both isolated impurities and those in *A* centers) are $9.48 \times 10^{20}$ cm$^{-3}$ and $4.0 \times 10^{19}$ cm$^{-3}$, respectively. The calculated $\varepsilon_f^{D-A}$ for CdTe:In and CdTe:Cl are $E_v$ + 1.23 eV and $E_v$ + 1.01 eV, respectively. Both of them are above the midgap. The calculated solid solubilities and $\varepsilon_f^{D-A}$ for In and Cl satisfy the requirements for obtaining high resistivity as discussed above. The lower $\varepsilon_f^{D-A}$ for CdTe:Cl than that for CdTe:In is due to the higher *A*-center binding energy calculated for Cl (1.01 eV) than for In (0.60 eV) (see Eq. 9).

The experimental results that high resistivity is maintained with [Cl] > $10^{19}$ cm$^{-3}$ suggests that the scenario depicted in Fig. 5 should apply to CdTe:Cl, which requires that $\varepsilon_f^{D-A}$ be confined within a narrow range (calculated to be from $E_v$ + 0.69 eV to $E_v$ + 0.88 eV for resistivity higher than $10^9$ Ωcm). Our calculated $\varepsilon_f^{D-A}$ for CdTe:Cl at the Te-rich limit ($E_v$ + 1.01 eV) is slightly higher than that range. ($\varepsilon_f^{D-A}$ for CdTe:Cl is the Fermi level where the formation energy lines of $Cl_{Te}$ and its *A* center ($V_{Cd}$- $Cl_{Te}$) intersect [see Fig. 7(a)].) However, the error in such magnitude (~0.2 eV) is not unreasonable in DFT calculations of defects. It may result from the errors in the formation energy of $V_{Cd}^{2-}$ and the *A* center binding energy ($\Delta B$) (see Eq. 9). Also, the comparison between the calculated and the measured resistivities are qualitative in nature, because the Fermi level in the real material is affected by many factors not reflected by the calculations, e.g, the nonequilibrium conditions during the crystal growth, defect kinetics during the crystal cool-down process, etc.

Under the Cd-rich and shallow-donor-rich conditions, both CdTe:In and CdTe:Cl should exhibit *n*-type conductivity as both $\varepsilon_f^{D-V}$ and $\varepsilon_f^{D-A}$ are near or even above the CBM [see Figs. 6(b) and 7(b)].

We have also calculated the formation energies of Cl and In induced DX centers in CdTe. Structures of these DX centers can be found in Ref. 57. For both donors, the shallow-donor and the DX-center energy lines intersect at a point above the CBM (see Figs. 6 and 7). Thus, the Cl and In DX centers are not stable unless the band gap can be increased by applying pressure or by alloying with Zn, as also found experimentally.[58, 59, 60] Our results show that the shallow donors are compensated by $V_{Cd}$ and *A* centers, not by DX centers. The experimental observation of the DX centers in $Cd_{0.8}Zn_{0.2}Te$ was made after annealing the sample in Cd vapor for several days, which eliminated both $V_{Cd}$ and *A* centers.[58, 59, 60]

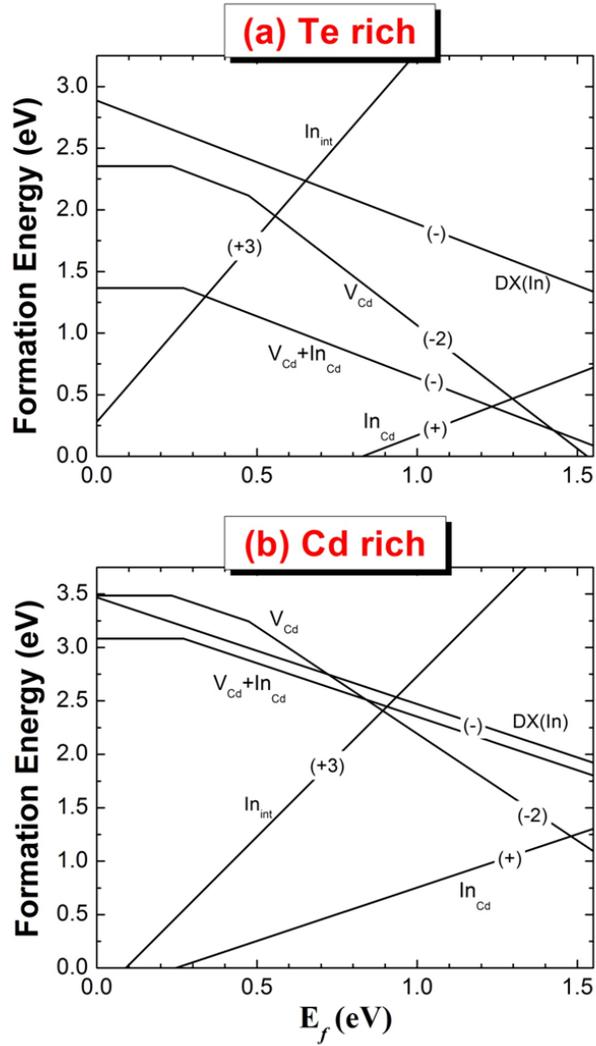

**FIGURE 6.** Formation energies of In-related defects, the Cd vacancy ($V_{Cd}^{2-}$), and the In-*A* center ($V_{Cd}^{2-}$-In$_{Cd}$) in CdTe at (a) Te-rich and (b) Cd-rich conditions. The slope of an energy line indicates the charge state of the defect, as selectively shown. The defect transition levels are given by the Fermi energy at which the slope changes.

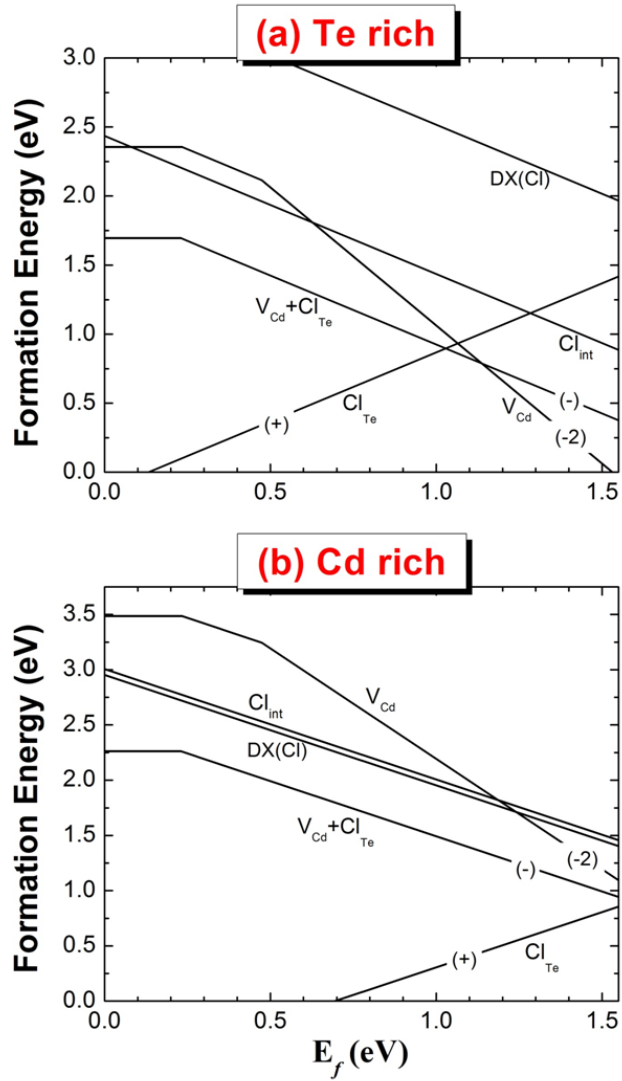

**FIGURE 7.** Formation energies of Cl-related defects, the Cd vacancy ($V_{Cd}^{2-}$), and the Cl-*A* center ($V_{Cd}^{2-}$-$Cl_{Te}$) in CdTe at (a) Te-rich and (b) Cd-rich conditions. The slope of an energy line indicates the charge state of the defect, as selectively shown. The defect transition levels are given by the Fermi energy at which the slope changes.

### D. Energetics of Native Defects

Figure 8 shows formation energies of important native defects in CdTe. Structures of these defects are similar to those reported previously.[27, 28] Comparing Fig. 8 to Figs. 6 and 7 shows that, under the In and Cl rich conditions, the formation energies of the substitutional In and Cl are lower than those of the native donor defects. Thus, with proper doping levels, the shallow donor, the Cd vacancy, and the *A* center can control the conductivity in CdTe. Compared to earlier LDA results,[27, 28] the hybrid functional calculations produce a lower VBM due to the partial correction of the self-interaction error. The consequence is that the deep defect levels in hybrid functional calculations are generally higher in energy relative to the VBM than in LDA calculations.

The low-energy native donors are $Cd_i$, $V_{Te}$, and $Te_{Cd}$, while $V_{Cd}$ acts as the dominant acceptor, as shown in Fig. 8. Using the results in Fig. 8, we obtain that $\varepsilon_f = E_v$ + 0.74 eV and $\rho = 1.2 \times 10^{10}$ cm$^{-3}$ at the Te-rich limit.[61] The calculated high resistivity for undoped CdTe seems to be inconsistent with the observation that the undoped CdTe is usually *p*-type with low resistivity. However, it is important to note that the high resistivity in undoped CdTe has also been obtained by using high-pressure Bridgman (HPB) technique,[5, 6] which suppresses the loss of Cd during the crystal growth. It has been pointed out by Szeles et al. that even the HPB grown CdTe suffers from some Cd loss and the incorporation of large concentration of Te precipitates and Te inclusions,[65] which indicates the Te-rich growth condition. It may be argued that the calculated defect energetics at the Te-rich limit is better compared to the HPB grown CdTe because the particle exchange with the gas phase is significantly reduced in the HPB growth. It may be difficult to reach thermal equilibrium when the bulk CdTe attempts to equilibrate with both Te precipitates and the Cd vapor simultaneously. The constant Te chemical potential assumed in the calculations may not reflect the reality during the crystal growth when non-equilibrium conditions exist. It is possible that the Cd vacancy concentration in CdTe grown by low-pressure techniques is higher than that when the thermal equilibrium is reached, and thus a low resistivity results. Even with HPB growth methods, the Cd loss during the crystal growth still exists. This may be related to the observation of large spatial variation of the resistivity in undoped CZT grown from HPB techniques,[62] which suggests that the non-equilibrium growth conditions play an important role in the carrier compensation in CdTe. However, the creation of additional Cd vacancies due to the non-equilibrium growth condition does not pose a problem for obtaining high resistivity in CdTe:Cl and CdTe:In since the solid solubilities of Cl and In in CdTe are sufficiently high to compensate the excess Cd vacancies.

The growth condition can be switched to Cd-rich by applying Cd overpressure. This will result in *n*-type CdTe as evidenced by the dominance of the low-energy donor defects shown in Fig. 8(b). Thus, controlling stoichiometry is the key to obtaining high resistivity in undoped CdTe. This, nevertheless, should be difficult, especially for obtaining uniform high resistivity in a large volume of the crystal. It is perhaps much easier to control the shallow donor concentration to be in the range of $10^{16}$ - $10^{17}$ cm$^{-3}$ than to control the native defect concentration to a similar range.

Fig. 6(b), 7(b), and 9(b) show that, under Cd-rich conditions, CdTe, regardless doped or undoped, should exhibit *n*-type conductivity. Many attempts have been made to remove Te-inclusions and precipitates by annealing CdTe and CZT under Cd

overpressure. This usually results in decrease in resistivity, [63, 64, 65, 66] consistent with our results.

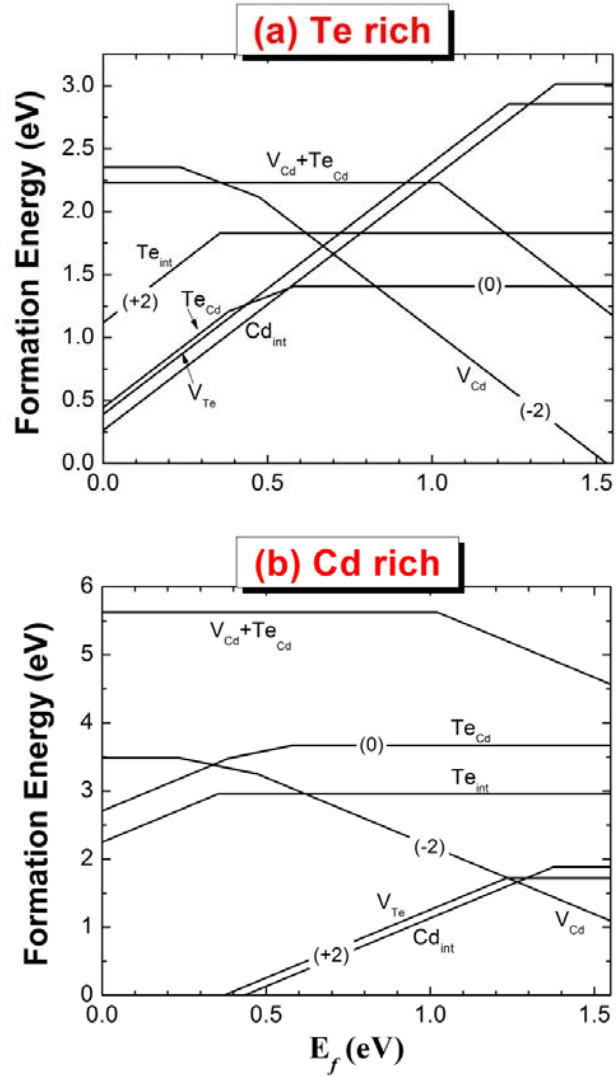

**FIGURE 8. Formation energies of native defects in CdTe at (a) Te-rich and (b) Cd-rich conditions. The slope of an energy line indicates the charge state of the defect, as selectively shown. The transition levels are given by the Fermi energy at which the slope changes.**

Figure 8 shows that $Cd_i^{2+}$ and $V_{Te}^{2+}$ have similar formation energies. However, our previous calculations show that $Cd_i^{2+}$ has a low diffusion barrier of 0.36 eV, [27, 28] which means that $Cd_i$ can be mobile at low temperatures. It is likely that most of $Cd_i^{2+}$ will recombine with $V_{Cd}^{2-}$ during the crystal cool-down process, which nevertheless does not affect the resistivity. $Te_{Cd}$ is a very deep donor with its (2+/+) and (+/0) levels calculated at $E_v$ + 0.38 eV and $E_v$ + 0.58 eV, respectively. The results in Fig. 8(a) shows that the Fermi level is pinned by $Te_{Cd}^+$ and $V_{Cd}^{2-}$. However, the calculated formation energies of $V_{Te}^{2+}$ and $Te_{Cd}^{2+}$ are very close. Removing the image charge correction will make $Te_{Cd}$ a negative-$U$ center and cause the Fermi level to be pinned by $V_{Te}^{2+}$ and $V_{Cd}^{2-}$. Despite this uncertainty, it is clear that $Te_{Cd}$ is an important donor when the growth condition is close to the Te-rich limit while $V_{Te}$ is important under both Te-rich and Cd-rich conditions (see Fig. 8). Regardless whether $Te_{Cd}$ or $V_{Te}$ is the more important donor under Te-rich conditions, the deep donor level of $Te_{Cd}$ will cause carrier trapping that is harmful to the detector performance. It is better to use shallow donor doping to pin the Fermi level at a somewhat higher energy than that determined by native defects, because this will significantly reduce the fraction of the ionized $Te_{Cd}$ that traps electrons.

$Te_{Cd}$-$V_{Cd}$ is a deep acceptor and is also a negative-$U$ center with its (0/2-) level located at $E_v$ + 1.02 eV (see Fig. 8). This defect may be related to the experimentally observed deep acceptor levels near the midgap.[25] But $Te_{Cd}$-$V_{Cd}$ has a high formation energy and thus should not play a significant role in carrier compensation.

## IV. Conclusions

We show that the carrier compensation between shallow donors (i.e., In, Cl) and acceptors (i.e., Cd vacancies, *A* centers) can explain the observed high resistivity and the associated shallow donor concentrations in CdTe. The carrier compensation mechanisms in Cl-doped, In-doped, and undoped CdTe may be summarized as follows: (1) In CdTe:Cl, the high resistivity is robust and may be obtained with a large variation of [Cl] ($10^{16}$ - $10^{19}$ cm$^{-3}$). The Fermi level is pinned near midgap by Cl$_{Te}$ and $V_{Cd}$ at relatively low [Cl] (e.g., $10^{16}$ cm$^{-3}$) and by Cl$_{Te}$ and the *A* center at relatively high [Cl] (e.g., $10^{19}$ cm$^{-3}$) [see Fig. 3(c) for an example]. The self-compensation between Cl$_{Te}$ and its *A* center ensures a fixed Fermi level when [Cl] is high, up to the Cl solubility. (2) In CdTe:In, high resistivity may be obtained with [In] ranging from $10^{16}$ - $10^{17}$ cm$^{-3}$. The Fermi level is pinned near midgap by In$_{Cd}$ and $V_{Cd}$ in semi-insulating CdTe. The [In] range of $10^{16}$ - $10^{17}$ cm$^{-3}$ is large enough to tolerate uncertainties in the doping level during the crystal growth. (3) In undoped CdTe, the high resistivity can only be achieved if the stoichiometry can be carefully controlled.

Among many room-temperature semiconductor radiation detector materials, the best electron µτ product has been reported to be better than $10^{-3}$ cm$^2$/V, e.g., in CdTe and TlBr.[67, 68] Our calculations on both CdTe and TlBr show that shallow donors, rather than deep ones, pin the Fermi level near midgap.[45] This makes considerable sense because the deep centers will reduce the electron µτ product significantly and thus cannot be present in large quantities in high-quality detector-grade CdTe and TlBr. These results suggest that, although deep centers are very effective in pinning the Fermi level and increase resistivity, one should seek proper shallow donors and shallow acceptors and optimize their concentrations to obtain good carrier compensation. The deep centers should be avoided in semiconductor radiation detectors.

## Acknowledgements

This work was supported by the U.S. DOE Office of Nonproliferation Research and Development NA22.

[1] T. E. Schlesinger, J. E. Toney, H. Yoon, et al., Mater. Sci. & Eng. **32**, 103 (2001)

[2] P. Rudolph, M. Hofmann, and M. Muhlberg, J. Cryst. Growth **128**, 582 (1993).

[3] P. Rudolph, Cryst. Res. Technol. **38**, 542 (2003).

[4] Cs. Szeles, Y. Y. Shan, K. G. Lynn, and E. E. Eissler, Nucl. Instrum. Methods Phys. Res. A **380**, 148 (1996).

[5] M. Fiederle, V. Babentsov, J. Franc, A. Fauler, J. –P. Konrath, Cryst. Res. Technol. **38**, 588 (2003).

[6] J. F. Butler, F. P. Doty, and B. Apotovsky, Mater. Sci. Eng. B**16**, 291 (1993).

[7] M. Fiederle, D. Ebling, C. Eiche, D. M. Hofmann, M. Salk, W. Stadler, K. W. Benz, and B. K. Meyer, J. Cryst. Growth **138**, 529 (1994).

[8] C. Eiche, D. Maier, D. Sinerius, J. Weese, K. W. Benz, and J. Honerkamp, J. Appl. Phys. **74**, 6667 (1993).

[9] Y. Xu, W. Jie, P. Sellin, T. Wang, W. Liu, G. Zha, P. Veeramani, and C. Mills, J. Phys. D: Appl. Phys. **42** 035105 (2009).

[10] M. Fiederle, A. Fauler, and A. Zwerger, IEEE Trans. Nucl. Sci. **54**, 769 (2007).

[11] M. Pavesi, L. Marchini, M. Zha, A. Zappettini, M. Zanichelli, and M. Manfredi, J. Elec. Mater. **40**, 2043 (2011).

[12] M. Fiederle, D. Ebling, C. Eiche, et al., J. Cryst. Growth **146**, 142 (1995).

[13] M. Fiederle, C. Eiche, M. Salk, R. Schwarz, K. W. Benz, W. Stadler, D. M. Hofman, and B. K. Meyer, J. Appl. Phys. **84**, 6689 (1998).

[14] G. F. Neumark, Phys. Rev. B **26**, 2250 (1982).

[15] V. Babentsov, J. Franc, H. Elhadidy, A. Fauler, M. Fiederle, and R. B. James, J. Mater. Res. **22**, 3249 (2007).

[16] M. Chu, S. Terterian, D. Ting, et al., Appl. Phys. Lett. **79**, 2728 (2001).

[17] N. Krsmanovic, K. G. Lynn, M. H., Weber, et al., Phys. Rev. B **62**, R16279 (2000).

[18] M. Fiederle, A. Fauler, J. Konrath, et al., IEEE Trans. Nucl. Sci. **51**, 1864 (2004).

[19] M. Fiederle, A. Fauler, J. Konrath, et al., IEEE Trans. Nucl. Sci. **51**, 1864 (2004).

[20] O. Panchuk, A. Savitskiy, P. Fochuk, Ye. Nykonyuk, O. Parfenyuk, L. Shcherbak, M. Ilashchuk, L. Yatsunyk, and P. Feychuk, J. Cryst. Growth **197**, 607 (1999).

[21] V. Babentsov, J. Franc, A. Fauler, M. Fiederle, and R. B. James, J. Mater. Res. **23**, 1751 (2008).

[22] Cs. Szeles, IEEE Trans. Nucl. Sci. **51**, 1242 (2004).

[23] V. Babentsov, J. Franc, and R. B. James, Appl. Phys. Lett. **94**, 052102 (2009).

[24] P. Fougeres, P. Siffert, M. Hageali, et al., Nucl. Instr. Methods Phys. Res. A **428**, 38 (1999).

[25] R. Soundararajan, K. G. Lynn, S. Awadallah, Cs. Szeles, and S. –H. Wei, J. Electr. Mater. **35**, 1333 (2006).

[26] V. Babentsov, V. Boiko, G. A. Schepelskii, R. B. James, J. Franc, J. Prochazka, and P. Hlidek, J. Luminescence **130**, 1425 (2010).

[27] M. –H. Du, H. Takenaka, D. J. Singh, Phys. Rev. B **77**, 094122 (2008).

[28] M. –H. Du, H. Takanaka, and D. J. Singh, J. Appl. Phys. **104**, 093521 (2008).

[29] S. A. Awadalla, A. W. Hunt, K. G. Lynn, H. Glass, C. Szeles, and S. –H Wei, Phys. Rev. B 69, 075210 (2004).

[30] S. –H. Wei and S. B. Zhang, Phys. Rev. B **66**, 155211 (2002).

[31] A. Carvalho, A. K. Tagantsev, S. Öberg, P. R. Briddon, and N. Setter, Phys. Rev. B **81**, 075215 (2010).

[32] J. P. Perdew, M. Ernzerhof, and K. Burke, J. Chem. Phys. **105**, 9982 (1996).

[33] J. Paier, M. Marsman, K. Hummer, and G. Kresse, I. C. Gerber, J. G. Angyan, J. Chem. Phys. **124**, 154709 (2006).

[34] P. Broqvist and A. Pasquarello, Appl. Phys. Lett. **89**, 262904 (2006).

[35] D. Muñoz Ramo, A. L. Shluger, J. L. Gavartin, and G. Bersuker, Phys. Rev. Lett. **99**, 155504 (2007).

[36] D. Muñoz Ramo, J. L. Gavartin, A. L. Shluger, and G. Bersuker, Phys. Rev. B **75**, 205336 (2007).

[37] A. V. Kimmel, P. V. Sushko, and A. L. Shluger, J. Non-Cryst. Solids **353**, 599 (2007).

[38] A. Alkauskas, P. Broqvist, and A. Pasquarello, Phys. Rev. Lett. **101**, 046405 (2008).

[39] A. Alkauskas, P. Broqvist, F. Devynck, and A. Pasquarello, Phys. Rev. Lett. **101**, 106802 (2008).

[40] F. Oba, A. Togo, I. Tanaka, J. Paier, and G. Kresse, Phys. Rev. B **77**, 245202 (2008).

[41] J. Paier, M. Marsman, and G. Kresse, Phys. Rev. B **78**, 121201 (2008).

[42] K. Hummer, J. Harl, G. Kresse, Phys. Rev. B **80**, 115205 (2009).

[43] M. –H. Du and S. B. Zhang, Phys. Rev. B **80**, 115217 (2009).

[44] A. Janotti, J. B. Varley, P. Rinke, N. Umezawa, G. Kresse, and C. G. Van de Walle, Phys. Rev. B 81, 085212 (2010).

[45] M. –H. Du, J. Appl. Phys. **108** 053506 (2010).

[46] M. –H. Du and K. Biswas, Phys. Rev. Lett. **106**, 115502 (2011).

[47] K. Biswas and M. –H. Du, Appl. Phys. Lett. **98**, 181913 (2011).

[48] H. –P. Komsa, P. Broqvist, and A. Pasquarello, Phys. Rev. B **81**, 205118 (2010).

[49] G. Kresse and J. Furthmüller, Phys. Rev. B **54**, 11169 (1996).

[50] *Landolt-Bornstein: Numerical Data and Functional Relationships in Science and Technology*, edited by O. Madelung, M. Schultz, and H. Weiss (Springer-Verlag, Berlin, 1982), Vol. 17b.

[51] G. Kresse and D. Joubert, Phys. Rev. B **59**, 1758 (1999).

[52] *CRC Handbook of Chemistry and Physics, 88th edition*, D. R. Lide, ed., CRC Press/Taylor and Francis, Boca Raton, FL.

[53] S. Lany and A. Zunger, Phys. Rev. B 78, 235104 (2008).

[54] J. P. Perdew, K. Burke, and M. Ernzerhof, Phys. Rev. Lett. **77**, 3865, (1996).

[55] *CRC Handbook of Chemistry and Physics, 88th edition*, D. R. Lide, ed., CRC Press/Taylor and Francis, Boca Raton, FL.

[56] K. Kim, J. Hong, and S. Kim, J. Cryst. Growth **310**, 91 (2008).

[57] M. –H. Du, Appl. Phys. Lett. **92**, 181908 (2008).

[58] T. Thio, J. W. Bennett, and P. Becla, Phys. Rev. B **54**, 1754 (1996).

[59] Y. Y. Shan, K. G. Lynn, Cs. Szeles, P. Asoka-Kumar, T. Thio, J. W. Bennett, C. B. Beling, S. Fung, and P. Becla, Phys. Rev. Lett. **79**, 4473 (1997).

[60] S. Fung, Y. Y. Shan, A. H. Deng, C. C. Ling, C. D. Beling, K. G. Lynn, J. Appl. Phys. **84**, 1889 (1998).

[61] The concentrations of the ionized and neutral Te$_{Cd}$ are calculated at high temperature of 1373 K without invoking the charge re-equilibration at the room temperature using Eq.8. The reason is that charging Te$_{Cd}$ causes significant electronic and structural relaxation. The ground-state structures of Te$_{Cd}^0$ and Te$_{Cd}^{2+}$ are of $C_{3v}$ and $T_d$ symmetries, respectively. The single-particle defect level of Te$_{Cd}^0$ is below the midgap while those of Te$_{Cd}^{2+}$ and Te$_{Cd}^+$ are much higher than the midgap. Therefore, the carrier statistics should not be affected much by reducing the temperature from the growth temperature to the room temperature.

[62] H. Yoon, M. S. Goorsky, B. A. Brunett, J. M. Van Scyoc, J. C. Lund, and R. B. James, J. Electr. Mat. **28**, 838 (1999).

[63] P. Rudolph, S. Kawasaki, S. Yamashita, S. Yamamoto, Y. Usuki, Y. Konagaya, S. Matada, and T. Fukuda, J. Cryst. Growth **161**, 28 (1996).

[64] J. Franc, R. Grill, P. Hlídek, E. Belas, L. Turjanska, P. Höschl, I. Turkevych, A. L. Toth, P. Moravec, and H. Sitter, Semicond. Sci. Technol. **16**, 514 (2001).

[65] C. Szeles, W. C. Chalmers, S. E. Cameron, J. –O. Ndap, M. Bliss, and K. G. Lynn, Proc. SPIE **4507**, 57 (2001).

[66] G. Young, A. E. Bolotnikov, P. M. Fochuk, G. S. Camarda, Y. Cui, A. Hossain, K. Kim, J. Horace, B. McCall, R. Gul, L. Xu, O. V. Kopach, and R. B. James, Proc. SPIE 780507 (2010).

[67] A. Owens and A. Peacock, Nucl. Instr. Method Phys. Res. A **531**, 18 (2004).

[68] A. V. Churilov, G. Ciampi, H. Kim, L. J. Cirignano, W. M. Higgins, F. Olschner, and K. S. Shah, IEEE Trans. Nucl. Sci. **56**, 1875 (2009).